\documentclass[conference]{IEEEtran}
\IEEEoverridecommandlockouts

\usepackage{cite}
\usepackage{amsmath,amssymb,amsfonts}
\usepackage{graphicx}
\usepackage{textcomp}
\usepackage[ruled]{algorithm2e}
\usepackage{algpseudocode}
\usepackage{xcolor}
\usepackage{braket}
\usepackage{multirow}
\usepackage{subfigure}
\usepackage{float}
\usepackage[switch,mathlines,displaymath]{lineno}

\usepackage[normalem]{ulem} 

\usepackage{etoolbox}
\makeatletter
\patchcmd{\@makecaption}
{\scshape}
{}
{}
{}
\makeatletter
\patchcmd{\@makecaption}
{\\}
{.\ }
{}
{}
\makeatother

\usepackage{caption}

\newcommand{\system}{Q$^{2}$Chemistry}

\newcommand{\SVD}{{\rm SVD}}

\newcommand{\change}{}

\def\BibTeX{{\rm B\kern-.05em{\sc i\kern-.025em b}\kern-.08em
    T\kern-.1667em\lower.7ex\hbox{E}\kern-.125emX}}

\begin{document}




\title{\change{Large-Scale Simulation of Quantum Computational Chemistry on a New Sunway Supercomputer}  \\


}

\author{\IEEEauthorblockN{Honghui Shang\IEEEauthorrefmark{2}\IEEEauthorrefmark{6}\IEEEauthorrefmark{1}, 
		Li Shen\IEEEauthorrefmark{3}\IEEEauthorrefmark{6},
		Yi Fan  \IEEEauthorrefmark{3},
		Zhiqian Xu\IEEEauthorrefmark{2},
		Chu Guo  \IEEEauthorrefmark{13}\IEEEauthorrefmark{14}\IEEEauthorrefmark{1},
		Jie Liu  \IEEEauthorrefmark{10}\IEEEauthorrefmark{1},  
		Wenhao Zhou \IEEEauthorrefmark{4}, \\
		Huan Ma \IEEEauthorrefmark{3},
		Rongfen Lin \IEEEauthorrefmark{11},
	    Yuling Yang \IEEEauthorrefmark{4}, 
		Fang Li \IEEEauthorrefmark{4}, 
		Zhuoya Wang \IEEEauthorrefmark{12},
		Yunquan Zhang\IEEEauthorrefmark{2},
		Zhenyu Li\IEEEauthorrefmark{3}
	}
	\IEEEauthorblockA{\IEEEauthorrefmark{2} Institute of Computing Technology, Chinese Academy of Sciences, Beijing, China}
	\IEEEauthorblockA{\IEEEauthorrefmark{3} University of Science and Technology of China, Hefei, China}
	
	\IEEEauthorblockA{\IEEEauthorrefmark{10} Hefei National Laboratory,  University of Science and Technology of China, Hefei, Anhui 230088, China}
	
	\IEEEauthorblockA{\IEEEauthorrefmark{13}
	Henan Key Laboratory of Quantum Information and Cryptography, Zhengzhou,
	Henan 450000, China} 
     
    \IEEEauthorblockA{\IEEEauthorrefmark{14}
     Hunan Normal University, Changsha 410081, China}

	\IEEEauthorblockA{\IEEEauthorrefmark{4}National Supercomputing Center in Wuxi, Wuxi, China}
	
	\IEEEauthorblockA{\IEEEauthorrefmark{11} Department of Computer Science and Technology, Tsinghua University, Beijing, China }
	
	\IEEEauthorblockA{\IEEEauthorrefmark{12} Pilot National Laboratory for Marine Science and Technology, Qingdao, China}

	\IEEEauthorblockA{\IEEEauthorrefmark{6}These authors contribute equally to this work; }
	\IEEEauthorblockA{\IEEEauthorrefmark{1}Corresponding authors:shanghonghui@ict.ac.cn; guochu604b@gmail.com;  liujie86@ustc.edu.cn}
}

\maketitle

\thispagestyle{plain}
\pagestyle{plain}

\begin{abstract}
\change{Quantum computational chemistry~(QCC) is the use of quantum computers to solve problems in computational quantum chemistry. We develop a high performance variational quantum eigensolver (VQE) simulator for simulating quantum computational chemistry problems on a new Sunway supercomputer.} The major innovations include: (1) a Matrix Product State (MPS) based VQE simulator to \change{reduce the amount of memory needed} and increase the simulation efficiency; (2) a combination of the Density Matrix Embedding Theory with the MPS-based VQE simulator to further extend the simulation range; (3) A three-level parallelization scheme to scale up to 20 million cores; (4) Usage of the Julia script language as the main programming language, which both \change{makes the programming easier and enables} cutting edge performance as native C or Fortran; (5) Study of real chemistry systems based on the VQE simulator, achieving nearly linearly strong and weak scaling.  Our simulation demonstrates the power of VQE for large quantum chemistry systems, thus paves the way for large-scale VQE experiments on near-term quantum computers.

\end{abstract}

\section{Introduction}
\change{Our world is fundamentally quantum mechanical. However, the application of the quantum mechanics to chemical problem “{\it leads to equations much too complicated to be soluble}”,  as Paul Dirac noted. Directly solving the many-electron Schr\"odinger equation, such as the full configuration interaction (FCI) method, has a complexity that grows exponentially with the problem size. Thus it would soon fail for moderate-size molecules with about $24$ electrons~(24 orbitals), which corresponds to a diagonalization problem of size $7.3$ trillion~\cite{VogMaOls17}.}

The advent of quantum computing technologies \change{allows a new approach} to solve quantum chemistry problems, referred to as quantum computational chemistry~\cite{CaoGuzik2019,McardleYuan2020}. 
In the long term, the quantum phase estimation algorithm can efficiently solve quantum chemistry problems with fault-tolerant quantum computers. In the short term, variational quantum eigensolver (VQE) provides a promising heuristic algorithm on noisy quantum computers~\cite{CerezoColes2021,BhartiGuzik2022}. VQE is believed to be friendly to near-term quantum computers in that it adapts to the qubit counts, the connectivity and the coherence time on noisy intermediate-scale quantum (NISQ) devices, while maintaining a shallow circuit to mitigate the effects of noise. 
Recently, quantum computational advantages for the problem of random circuit sampling have been demonstrated on noisy quantum computers with between $50$ and $100$ qubits~\cite{AruteMartinis2019,WuPan2021,ZhuPan2022}. However, demonstrating practical quantum advantages on real world problems such as quantum chemistry problems \change{remains a great challenge for noisy quantum computing}, for which VQE based quantum chemistry solver is a promising candidate. 
A variety of electronic structure problems to achieve an industrially relevant computational advantage on contemporary and near-future quantum computers have been discussed recently~\cite{ElfBroWeb20,BauBraMot20}. 

At the current stage, classical simulation of quantum computation is crucial for the study of quantum algorithms and quantum computing architectures, especially for heuristic quantum algorithms as VQE. \change{The largest VQE experiment performed on a quantum computer up to date has used $12$ qubits~\cite{Google2020a}.}  An industrially relevant quantum computational advantage in quantum chemistry is expected to appear at around $38\leq N \leq 68$ qubits (under the assumption of error-corrected qubits), which is related to an electronic structure problem including $19\leq N \leq 34$ electrons~\cite{ElfBroWeb20}. However, as a heuristic algorithm, it is \change{\textit{a priori}} not clear that the given ansatz for the parametric quantum circuit is expressive enough to represent the ground state of the problem and that even it is, whether the iterative algorithm can converge to the ground state. 
Moreover, current quantum computers are noisy and the errors of quantum gate operations are often dependent on the types of the gates as well as the qubits that they act on.
As a result the accuracy and performance scaling of VQE with the problem size and the complexity of the parametric circuit ansatz may be hardware specific and also problem specific.
Therefore before running a full fledged VQE application on current quantum computers with possibly hundreds of qubits, the accuracy and performance scaling, robustness as well as limitations of VQE should first be explored using a classical simulator. This situation is similar to the random quantum circuit (RQC) sampling task used to demonstrate quantum supremacy, where a benchmark baseline from noiseless classical simulation is in need to fully characterize its precision~\cite{GuoWu2019,VillalongaMandra2019,GuoHuang2021,PanZhang2021,PanZhang2021b,HuangChen2021,LiuChen2021}. Compared to RQC, VQE is much more demanding for both quantum and classical computers, for example, the number of CNOT gates involved in a typical \change{quantum computational chemistry} simulation quickly goes beyond $10^6$ with commonly used physically motivated ansatz such as unitary coupled-cluster (UCC). Moreover, the parametric quantum circuit has to be executed many times as is typical for variational algorithms. These effects limit most of the current investigations of VQE using classical computers to very small problems (less than $20$ qubits).




In view of these challenges for practical application of VQE in the age of noisy quantum computing, we believe a highly efficient and large-scale classical simulator of VQE on a leading supercomputer is currently in need, in that, 1) it allows to analyze the scaling of VQE for practical quantum chemistry systems; 2) it allows a step by step cross verification of near-term VQE algorithms running on noisy quantum computers with around $100$ qubits, which is essential, for example, to quantify the quantum computational advantage; 3) this algorithm itself is also a highly parallelizable classical algorithm to solve large-scale quantum chemistry problems and thus would have practical importance on its own.

Based on the above considerations, we develop a highly efficient VQE simulator for the new generation Sunway supercomputer. We use the physically motivated UCC ansatz for the parametric quantum circuit of VQE~\cite{UCCansatz}. In view of the nearest-neighbour nature of such parametric quantum circuit, we use the Matrix Product State (MPS) to efficiently represent the underlying quantum state~\cite{Schollwock2011}. Our MPS-VQE simulator can efficiently simulate around $100$-qubit quantum circuit with moderate circuit depth. Since the majority of real-world molecules have a very large number of active electrons, which would often require more than $1000$ qubits in VQE, we further extend our capability of our MPS-VQE simulator using the Density Matrix Embedding Theory (DMET)~\cite{LiLv2021}. The combination of these techniques allow us to address quantum chemistry problems with up to $1000$ qubits with descent accuracy.
The major innovations of the current work are summarized as follows: 
\begin{itemize}
	\item The MPS representation of the quantum state is used to explore the low-entanglement nature for relatively shallow quantum circuits, which  \change{significantly reduces} the memory requirement and the computation time for VQE \change{(the largest VQE simulation in this work uses $200$ qubits)}. 
	\item The divide-and-conquer technique for quantum many-body systems, DMET, is used in combination with the MPS-VQE simulator to further extend the simulation scale.
	\item A three-level parallelization scheme from DMET to MPS and then down to elementary tensor operations to scale up to $20$ million cores. 
	\item The Julia script language is used as the main programing language, which allows to achieve high performance without the need to perform any low-level operations using systems languages such as C and Fortran. The fast prototyping nature of the Julia language as well as its rich scientific libraries and benchmarking tools make optimizations much easier.
	\item Real chemistry systems have been studied, and nearly linearly strong and weak scaling have been achieved. 
\end{itemize}

These innovations enable us to perform quantum computational chemistry simulations on the scale of hundreds of atoms. \change{It should be noted that although the ingredients used in this work, namely MPS, DMET, VQE, Julia, exist before this work, a full integration of those methods to simulate the realistic large chemical systems is nontrivial and has not been done before. 
For example, MPS has not been combined with DMET or VQE before.
A highly parallelizable design of the simulation scheme and an efficient implementation of it to scale up to over 20 million cores is also original in this work. Beside, unleashing the Julia programming language on Sunway architectures and running it efficiently over 20 million cores is also an extremely challenging task. Briefly speaking, our work has set the standard for large-scale classical simulation of quantum computational chemistry, and  paves the way for benchmarking VQE applications on near-term noisy quantum computers.}

\section{Background}

\subsection{Current state of the art}

VQE approximates the exact eigenstate with a unitary transformation acting on a reference state. The reference state is often chosen to be some state that can be easily prepared on a quantum computer and the unitary transformation is represented by a parametric quantum circuit, with the parameters updated iteratively based on the output energy of each minimization step. A classical simulator for VQE simulates this process on a classical computer. The brute-force approach to simulate a quantum circuit is the state vector simulator, which exactly stores the underlying $n$-qubit quantum state.
The memory cost of the state vector simulator scales as $2^n$ for pure quantum states (implemented in {\sc ProjectQ },~\cite{projectq} {\sc Qiskit},~\cite{qiskit}  {\sc QuEST}~\cite{Quest} and {\sc Qulacs}~\cite{Qulacs}) and $2^{2n}$ for mixed quantum states (implemented in {\sc DM-Sim }, {\sc Qiskit},~\cite{qiskit}  {\sc QuEST}~\cite{Quest} and {\sc Qulacs}~\cite{Qulacs}). The largest quantum circuit simulation with this approach has reached $45$ qubits, but it would be extremely difficult to go further beyond due to the exponential memory cost~\cite{HanerSteiger2017}.

Tensor network based algorithms are important alternatives that reduce the memory requirement by representing the quantum state as a tensor network of low-rank tensors, which have been widely applied to simulate RQCs~\cite{GuoWu2019,VillalongaMandra2019,GuoHuang2021,PanZhang2021,PanZhang2021b,HuangChen2021,LiuChen2021}. The tensor network methods used for RQCs are mostly based on high-dimensional tensor network contractions since the RQCs under study usually have well defined two-dimensional geometries. However, the parametric quantum circuits used in VQE would not have such clear geometrical features in general, especially for those physical motivated ansatz such as unitary coupled-cluster as used in this work. In such instances the MPS (a very special type of one-dimensional tensor network that can efficiently simulate weakly entangled quantum states) based simulator is often preferable, especially for quantum chemistry problems. We also note that MPS has already been widely applied to solve quantum chemistry problems by directly computing the ground states in the context of Density Matrix Renormalization Group (DMRG) algorithm~\cite{WhiteMartin1999,ChanSharma2011}. MPS has also been used to directly simulate Shor's algorithm up to $60$ qubits~\cite{DangHollenberg2018}. MPS-based simulator has already been implemented, for example in QISKIT. However, large-scale implementation of it for VQE on a top supercomputer has not been demonstrated.



The electronic Hamiltonian of a chemical system can be written in a second-quantized formulation:
	\begin{equation}
		\begin{aligned}
			\hat{H}&=\sum_{p,q}{h_{q}^{p}a_{p}^{\dagger}a_{q}} +\frac{1}{2}{\sum_{\substack{p,q\\r,s}}}{g_{rs}^{pq}a_{p}^{\dagger}a_{q}^{\dagger}a_{r}a_{s}}
		\end{aligned}
		\label{eq:ham-pbc}
	\end{equation}
	where $h_{q}^{p}$ and $g_{rs}^{pq}$ are one- and two-electron integrals in molecular orbital basis.
	In a VQE procedure, the creation and annihilation operators $\{{a_{p}^{\dagger}}, a_{q}\}$ in the Hamiltonian should be mapped to weighted Pauli strings using the Jordan-Wigner or Bravyi-Kitaev transformation, therefore the energy can be obtained by measuring and summing the expectation values of Pauli strings $\{P\}\in\{I, \sigma_{x},\sigma_{y},\sigma_{z}\}^{\otimes n} $ as
	\begin{equation}
		\begin{aligned}
			E&=\langle{\hat{H}}\rangle=\langle \Psi(\theta)|\hat{H}|\Psi(\theta) \rangle\\
			&=\langle\Psi(\theta)|\sum_{p, q}{\tilde{h}_{q}^{p} P_{q}^{p}}|\Psi(\theta) \rangle \\
			&+ \langle \Psi(\theta)| \sum_{p,q,r,s}{\tilde{g}_{rs}^{pq}P_{rs}^{pq}}|\Psi(\theta) \rangle.
		\end{aligned}
		\label{eq:qubit-ham}
	\end{equation}
	The accuracy of VQE heavily depends on the wave function ansatz (or the structure of the parametric quantum circuit) used to approximate the unknown ground state of the problem Hamiltonian.
The mostly used form of the wave function ansatz is the physically motivated unitary coupled-cluster (UCC)\cite{UCCansatz}, where the wave function takes the form
\begin{equation}
	|\Psi(\theta)\rangle=e^{\hat{T}(\theta)-\hat{T}^{\dagger}(\theta)}|\Phi_{0}\rangle.
\end{equation}
Here $|\Phi_{0}\rangle$ is an initial state in the form of a single Slater determinant. When truncated to the single and double excitations (UCCSD), the cluster operators are
\begin{equation}
	\hat{T}(\theta)=\sum_{a,i}\theta_{i}^{a}\hat{a}^\dag_{a} \hat{a}_{i} +\frac{1}{4}{\sum_{a,b,i,j}} \theta_{i j}^{a b} \hat{a}_{a}^{\dagger}\hat{a}_{b}^{\dagger} \hat{a}_{i} \hat{a}_{j},
\end{equation}
where $\{i, j, k, \dots \}$, $\{a, b, c, \dots\}$ and $\{p, q, r, \dots\}$ indicate the occupied, virtual, and general orbitals, respectively. The UCCSD method has no exact finite truncation for the Baker-Campbell-Hausdorff expansion, which indicates that it can not be effectively implemented on a classical computer using polynomial expansion methods or the explicit matrix-vector method. UCCSD can be implemented as a parametric quantum circuit by Suzuki-Trotter decomposition of the unitary evolution operator into one- and two-qubit gates~\cite{PouHasWec15}.

Many contributions have been devoted to reducing the complexity of the quantum circuit ansatz for quantum chemistry problems. Recently, the excited states of Ir(ppy)$_3$ have been computed with up to $56$ qubits \change{\cite{RyaIZmGen21}}. This is the largest classical simulation of VQE for quantum computational chemistry in terms of the number of qubits up to date. However, to achieve this simulation a very shallow parametric quantum circuit ansatz (with around $60$ two-qubit gates) was employed to reduce the computational complexity. 


Following the philosophy that ''{\it divide each of the difficulties under examination into as many parts as possible, and as might be necessary for its adequate solution}'', a general VQE framework combined with the divide-and-conquer method (DMET)~\cite{Knizia2012} for large-scale simulation of quantum chemistry has been proposed~\cite{YamMatNar18}. Using this approach, the C$_{18}$ molecule with $144$ qubits is studied classically, where the VQE in each fragment uses at most $16$ qubits~\cite{LiHuaCao21}.
A similar idea has been adopted in a hybrid tensor network approach, in which a bottom-layer tensor network is introduced to describe the strongly correlated intrasubsystem interactions and a top-layer tensor network is used to describe the intersubsystem interactions~\cite{YuaSunLiu21}. This embedding method can also be applied to excited state calculation~\cite{Simon2017,Qiao2021}. 

Overall, the current state-of-the-art quantum computational chemistry simulators for large-scale systems are limited with respect to computational efficiency and scalability. Thus, the classical simulation of large-scale quantum chemistry systems are severely limited. This represents the key bottleneck in simulating the physical properties of real systems.

The goal we want to achieve is to faithfully simulate the large-scale VQE experiments for practical Chemical systems running on quantum computers, using a world-leading supercomputer. Since we are aiming for a parallelization scale of up to tens of millions of cores, the algorithm-wise challenge is to identify a suitable algorithm which could map the computation onto all the available cores of the underlying supercomputer, thus extending the simulation range of VQE to quantum chemistry systems which are of practical relevance.

The state vector simulator is out of choice due to the huge memory requirement. We thus take the tensor network based simulator which reduces the exponential memory growth. In practice we find that MPS based simulator gives the best performance when used in combination with the UCC ansatz. To spread the computations of the MPS-VQE simulator over the new Sunway supercomputer, there are three implementation-wise challenges:

\begin{itemize}
\item To parallelize over the $20$ million cores, we need a scheme which can first efficiently decompose the VQE simulation into sub-tasks (independent quantum circuit evolution and measurements) with little communication overhead and of similar complexity, and then distribute the tensor operations involved in a single circuit evolution in each process.
\item To achieve maximal efficiency for the major tensor operations, which are SVD and tensor contraction, we need a well tuned map of these tensor operations to the underlying CPE mesh.
\item Given the highly complex nature of the program including a hybridization of DMET, MPS and VQE as well as the application for practical problems, a variety of scientific libraries in high-level programming languages such as Python needs to be used to avoid reinventing the wheels. To build such a complex application with the simplest coding and also maintain a computational efficiency close to the low level C language, we pay a huge effort to develop the Julia compilation framework with SACA runtime support on heterogeneous processors to achieve functional implementation and performance improvement. 
\end{itemize}

\subsection{HPC System and Environment}

The new-generation Sunway supercomputer is used for performance 
assessment in this work, which is the successor of the Sunway
TaihuLight supercomputer. Similar to the Sunway TaihuLight system, the new Sunway supercomputer adopts a new generation of domestic  high-performance heterogeneous many-core processors~(SW26010Pro) and interconnection network chips in China. 

\begin{figure}
	\centerline{\includegraphics[width=0.93
		\columnwidth]{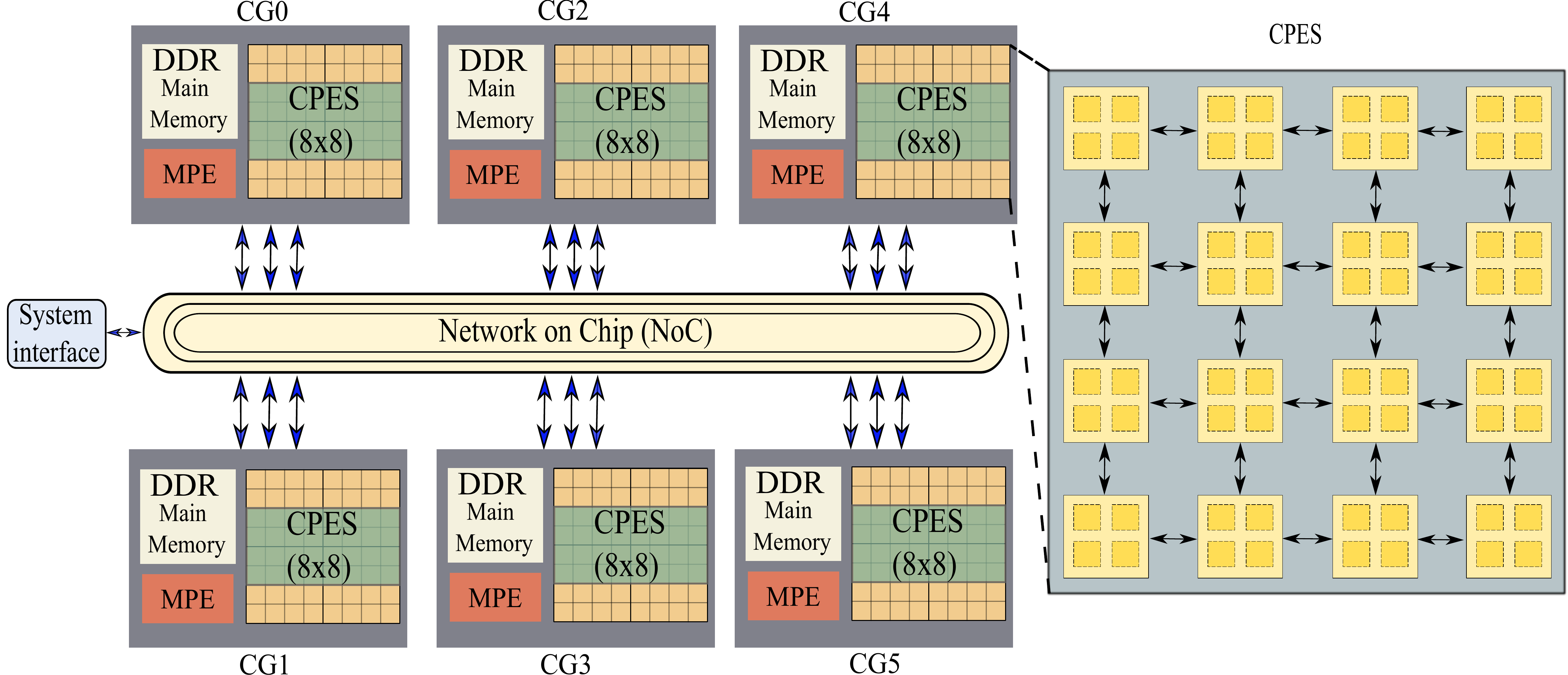}}
	\caption{The architecture of the SW26010Pro processor.}
	\label{fig:swcpu}
\end{figure}

The new SW processor is designed for massive thread and data parallelism and to deliver high performance on parallel workloads.  
The architecture of the SW26010Pro processor is shown in Fig.~\ref{fig:swcpu}.  Each processor contains 6 core-groups (CGs), with 65 cores in each CG, and in total 390 cores. Each CG contains one management processing element (MPE), one cluster of computing processing elements (CPEs) and one memory controller (MC). The MPE within each CG is 
used for computations, management and communication. The CPEs are organized as an $8\times 8$ mesh~(64 cores) and are designed to maximize the aggregated computing power and to minimize the complexity of the micro-architecture.
The CPEs are organized with a mesh network to achieve high-bandwidth data communication (P2P and collective communications) among the CPEs in one CG, which is referred to as the remote scratchpad memory access~(RMA). 

Each SW26010Pro processor contains 96 GB memory, with 16 GB memory in each CG. The MPE and the CPEs within the same CG share the same memory which is controlled by the MC. 
Each CPE has a 32 KB L1 instruction cache, and a 256 KB  scratch pad memory (SPM, also called the Local Data Memory (LDM)), which serves the same functionality as the L1 cache.  The data storage space can also be configured as a local data cache (LDCache) which is automatically managed by the hardware. Data transfer between LDM and main memory can be realized by direct memory access (DMA), and data transfer between LDCache and main memory can also be realized by conventional load/storage instructions. 

\section{Innovations}
\label{sec:innovations}
\subsection{The MPS-VQE simulator}
\begin{figure}
	\centerline{\includegraphics[width=0.93\columnwidth]{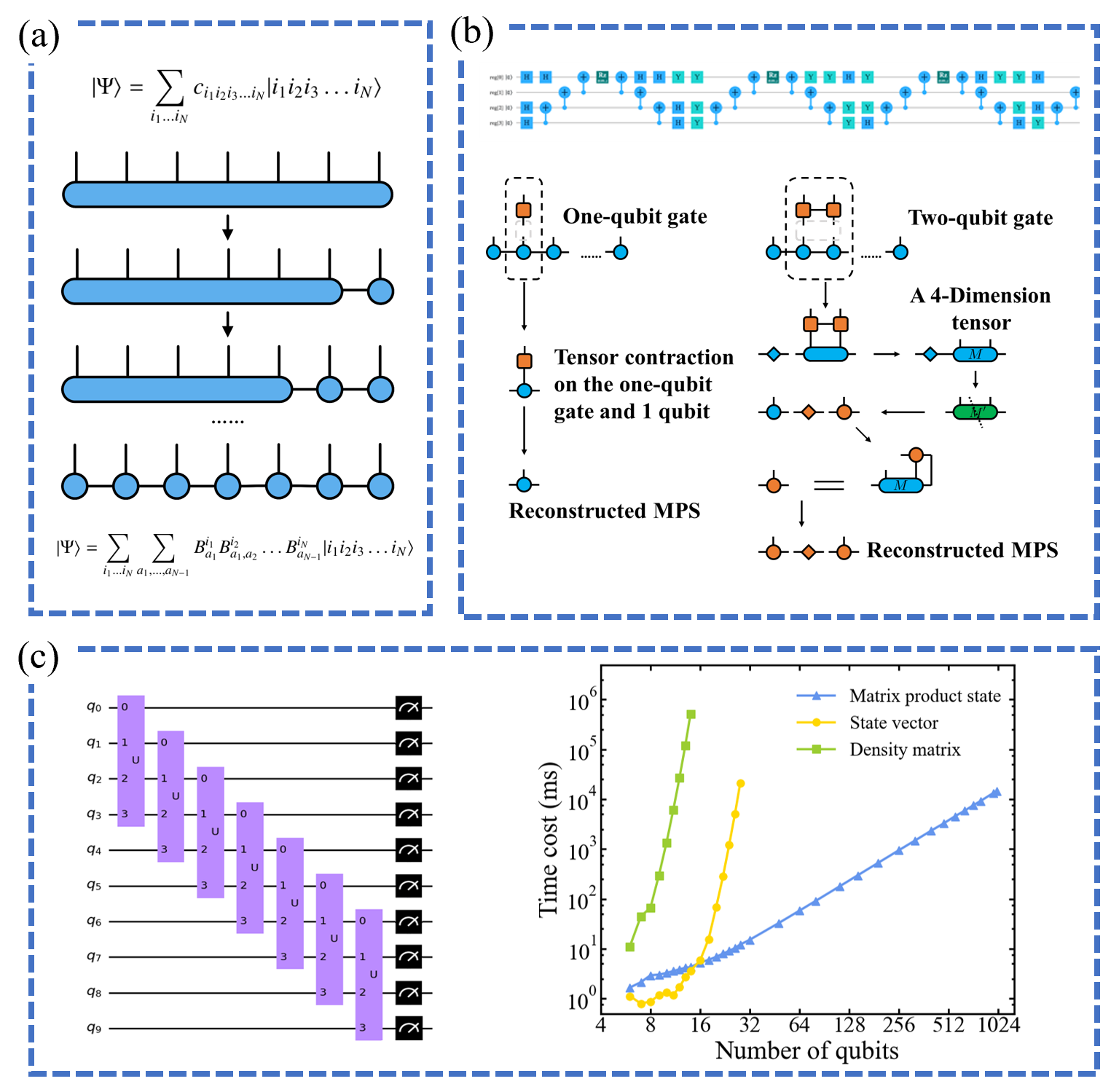}}
	\caption{(a) Decomposition of the FCI tensor into an MPS. (b) Application of single-qubit and two-qubit quantum gate operations onto the MPS. (c) The performance scaling with the number of qubits for the state vector~(SV), density matrix~(DM) and matrix product state~(MPS) simulators for a  quantum circuit \change{which entangles every $4$ consecutive qubits}. }
	\label{fig:mps}
\end{figure}


In quantum chemistry, the FCI generates a correlated wave function by considering all the excitations above the reference Hartree-Fock state. The FCI wave function $\left|\Psi\right\rangle$ can be written as:
\begin{equation}\label{eq:wavefunction}
	\left|\Psi\right\rangle  = \sum_{i_1...i_N} c_{i_1 i_2 i_3 \ldots i_N} |i_1 i_2 i_3 \ldots i_N \rangle
\end{equation} 
where $|i_1 i_2 i_3 \ldots i_N \rangle$ is the computation basis used, the coefficient $c_{i_1 i_2 i_3 \ldots i_N}$ is a rank-$N$ tensor of $2^N$ complex numbers.  
Finding the tensor $c$ corresponding to the ground state is in generally exponentially hard. Various simplified ansatz has been used instead of the exact form in Eq.(\ref{eq:wavefunction}) to reduce the exponential scaling, among which the tensor network states (TNSs) constitute an outstanding class~\cite{Schollwock2011,Orus2014}. TNS represents the underlying high-rank tensor as the product of a tensor network made of low-rank tensors, and the number of parameters in a TNS typically scales linearly with the system size (number of qubits). MPS is a special class of one-dimensional TNS which not only allows efficient representation, but also enables efficiently manipulations, such as the local gate operations and evaluating local observables, on it. In the context of random circuit sampling, it has been shown that the two-dimensional variant of TNS, namely the Projected Entangled Pair States (PEPS), would have better performance than MPS since current random circuits are mostly two-dimensional~\cite{GuoWu2019}. Here we will use the MPS ansatz for the quantum state due to the complex geometry of our parametric circuit ansatz, as shown in Fig.~\ref{fig:mps}(a), which can be written as
\begin{equation}\label{eq:mps}
	c_{i_1 i_2 \ldots i_N} = \sum_{a_1, \dots, a_{N-1}} B^{i_1}_{a_1} B^{i_2}_{a_1, a_2} \dots B^{i_N}_{a_{N-1}},
\end{equation}
with $i_n$ the `physical' index and $a_n$ the `auxiliary' index. The largest size of the auxiliary indices is referred to as the \textit{bond dimension} of MPS, denoted as $D = \dim_{1\leq n<N}(a_n)$. Eq.(\ref{eq:mps}) can represent arbitrary quantum states if $D$ is allowed to grow exponentially. A quantum state is said to be efficiently representable as an MPS if $D$ remains almost constant when $N$ grows. An MPS based algorithm in general has a complexity of $O(D^3)$. In practice, it is often advantageous to keep MPS in a canonical form, for example, the right-canonical form by ensuring that $\sum_{i_n, a_n} (B^{i_n}_{a_{n-1}, a_n})^{\ast} B^{i_n}_{a_{n-1}', a_n} = \delta_{a_{n-1}, a_{n-1}'} $ (for one thing, the algorithm would be numerically more stable since the tensors on each site are well normalized).

VQE requires to apply single-qubit and two-qubit gate operations sequentially onto the quantum state and then compute the expectation value of a (local) observable on the quantum state, which is similar to the time-evolving block decimation algorithm used for time evolution of one-dimensional Hamiltonians~\cite{Vidal2003}. Assuming that the initial MPS is prepared in the right-canonical form, a nearest-neighbour two-qubit gate operation, denoted as $O^{i_n, i_{n+1}}_{i_{n}', i_{n+1}'}$, can be applied onto a nearest-pair of tensors on sites $n$ and $n+1$ (also referred to as the $n$-th bond) as follows~\cite{Hastings2009} (The implementation of single-qubit gate operation is straightforward, but it is not necessary since single-qubits gate can be absorbed into two-qubits gate using gate fusion). As shown in Fig.~\ref{fig:mps}(b), first, one obtains the rank-$4$ tensor $M$ by the contraction of the two-qubit gate with the tensors $B^{i_n}_{a_{n-1}, a_n}$ and $B^{i_{n+1}}_{a_{n}, a_{n+1}}$
\begin{equation}
	M_{a_{n-1}, i_n, i_{n+1}, a_{n+1}} = \sum_{a_n} O^{i_n, i_{n+1}}_{i_{n}', _{n+1}'}  B^{i_n'}_{a_{n-1}, a_n} B^{i_{n+1}'}_{a_{n}, a_{n+1}}.
\end{equation}
Now to restore the MPS form, one could perform a singular value decomposition (SVD) on $M$ to obtain two rank-$3$ site tensors, however one of them will no longer be right-canonical. To preserve the right canonical form, one could perform SVD on another rank-$4$ tensor 
\begin{equation}
	M'_{a_{n-1}, i_n, i_{n+1}, a_{n+1}} = \sum_{a_{n-1}} \lambda_{a_{n-1}} M_{a_{n-1}, i_n, i_{n+1}, a_{n+1}}
\end{equation}
instead, where $\lambda_{a_{n-1}}$ are the Schmidt numbers on the $n-1$-th bond saved from the previous steps, that is,
\begin{equation}
	\SVD(M'_{a_{n-1}, i_n, i_{n+1}, a_{n+1}}) = \sum_{a_n} U^{i_n}_{a_{n-1}, a_n} \lambda_{a_n}' V^{i_{n+1}}_{a_n, a_{n+1}}, 
\end{equation}
where $\lambda_{a_n}'$ are the new Schmidt numbers on the $n$-th bond which will be used to replace the previous ones. In case the number of nonzero Schmidt numbers is larger than $D$, we will also truncate them back to $D$ by only reserving the $D$ largest Schmidt numbers. This procedure will produce truncation errors if the maximum $D$ we choose is too small or the circuit is too deep (Nevertheless this error can be monitored as an indication of the imprecision of the computing). $V^{i_{n+1}}_{a_n, a_{n+1}}$ is already right-canonical and will be used to replace $B^{i_{n+1}}_{a_n, a_{n+1}}$, the new site tensor on site $n$ can be updated as
\begin{equation}
	B^{i_n}_{a_{n-1}, a_n} = \sum_{i_{n+1}, a_{n+1}} M_{a_{n-1}, i_n, i_{n+1}, a_{n+1}} (V^{i_{n+1}}_{a_n, a_{n+1}} )^{\ast}
\end{equation}
which can be verified to also be right-canonical.  
With a right-canonical form of MPS, the expectation value of a local observable $\hat{O}_n$ (with its matrix representation $O_{i_n'}^{i_n}$ in the computational basis) on the MPS can be evaluated as 
\begin{equation}
	\langle \Psi \vert \hat{O} \vert \Psi\rangle = \sum_{a_{n-1}, a_n, i_n, i_n'} \lambda_{a_{n-1}}^2 O_{i_n'}^{i_n} B^{i_n'}_{a_{n-1}, a_n} (B^{i_n}_{a_{n-1}, a_n} )^{\ast}.
\end{equation}
The expectation value of two-qubit or higher observables can be efficiently computed accordingly. In this work, the expectation values of Pauli strings are calculated using Hadamard test, which is equivalent to evaluating the expectation value of the local projector $|0\rangle\langle0|$ or $|1\rangle\langle1|$ on the ancillary qubit.

To this end, we note that since our VQE is based on the MPS representation, the expressiveness of the VQE ansatz used in our classical simulator will not exceed the expressiveness of the underlying MPS. As a result one may well substitute the VQE simulator by another MPS based optimization algorithm such as DMRG and a similar or even higher precision would be expected if the same $D$ is used. However, the MPS based VQE simulator, as will be shown later, allows straightforward parallelization to millions of cores with almost no data communications, in comparison in current parallel algorithms for DMRG data communication can not be avoided and the parallelizability is limited~\cite{StoudenmireWhite2013}.

The MPS based simulator takes advantage of the low entanglement nature of the underlying quantum state to dramatically reduce the required computational resources, which is often true for NISQ devices for which decoherence limits the amount of entanglement that can be generated (except for RQC which quickly entangles the quantum states but usually has no clear physical meaning). An illustrative comparison between the performances of state vector~(SV), density matrix~(DM) and MPS simulators is shown in Fig.\ref{fig:mps}(c), \change{with a quantum circuit that generates a quantum state which can be written as an MPS with $D=8$. This circuit implements unitaries each of which entangles 4 consecutive qubits, corresponding to the correlations between neighbouring orbitals in a chemical system. It is also used as the MPS-inspired ansatz\cite{LiuZhang2019mps, Jack2022tn} for slightly correlated molecules.}  For the SV or DM method, the exponentially growing memory storage and computation complexity strictly limited the simulation scale, while the MPS method approximates the low-entangled quantum states with smaller tensors, that only spans a very tiny fraction of the overall Hilbert space, so the required computational resources is dramatically reduced. 
However, faithfully simulating of the whole VQE calculation is still numerically demanding since a huge number of quantum circuits~($\mathcal{O}(N_{q}^{4})$ for $N_{q}$) has to be evaluated. Nevertheless, this is an ideal situation for massive parallelization, which will be discussed in  Sec.~\ref{subsection:para}.

\subsection{The combination of DMET method with MPS-VQE Simulator}

\begin{figure}
	\centerline{\includegraphics[width=0.98\columnwidth]{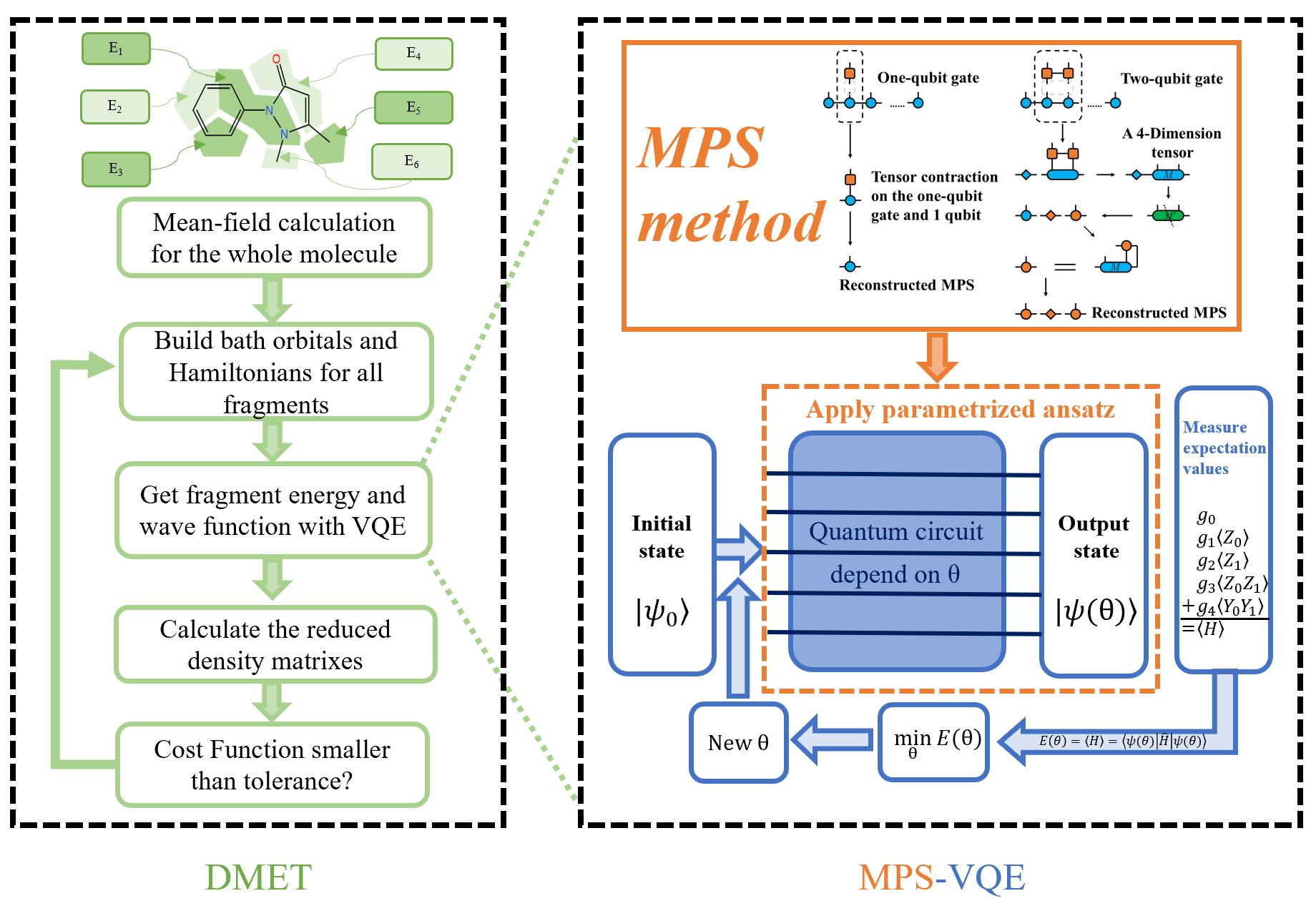}}
	\caption{Flowchart of our quantum computational chemistry simulator, the Julia version of \system. }
	\label{fig:flowchart}
\end{figure}

To facilitate the quantum simulation of the large systems, we use density matrix embedding theory (DMET) to reduce large systems into smaller fragments and use MPS-VQE as a fragment solver, as shown in Fig.~\ref{fig:flowchart}. 
The total energy of the system is the sum of the energies of the fragments,  each of which is embedded in an effective environment (quantum bath) compressed from the rest of the whole system. DMET provides an economic and highly accurate way to describe the interaction between the fragment and the effective environment. In this way, the electronic structure of the system can be computed by iteratively solving a bunch of smaller problems, which can be done using the state vector or MPS simulators (or ultimately using a quantum computer).
In practice, the standard procedures of DMET-MPS-VQE method are listed as follows:
\begin{enumerate}
	\item Perform low-level calculation for the whole system.
	\item Divide the whole system into fragments with reasonable number of orbitals for each fragment.
	\item For every fragment, construct bath orbitals and the reduced Hamiltonian by Schmidt decomposition and projection. 
	\item Get the fragment energy and the wave function with VQE, and then compute the number of electrons with 1-RDM~(reduced density matrix).
	\item Check if the sum of the number of electrons in the fragments agrees with the number of electrons for the entire system. If not, modify the global chemical potential and go back the step 3.
\end{enumerate}

In DMET, a high-level calculation for each fragment is carried out individually until the self-consistency criterion has been met: the sum of the number of electrons of all of the fragments~\change{(the number of fragments are around 10 to 200)} agrees with the number of electrons for the entire system.

\subsection{The Customized Parallelization Strategy}
\label{subsection:para}
\begin{figure*}
	\centerline{\includegraphics[width=1.8\columnwidth]{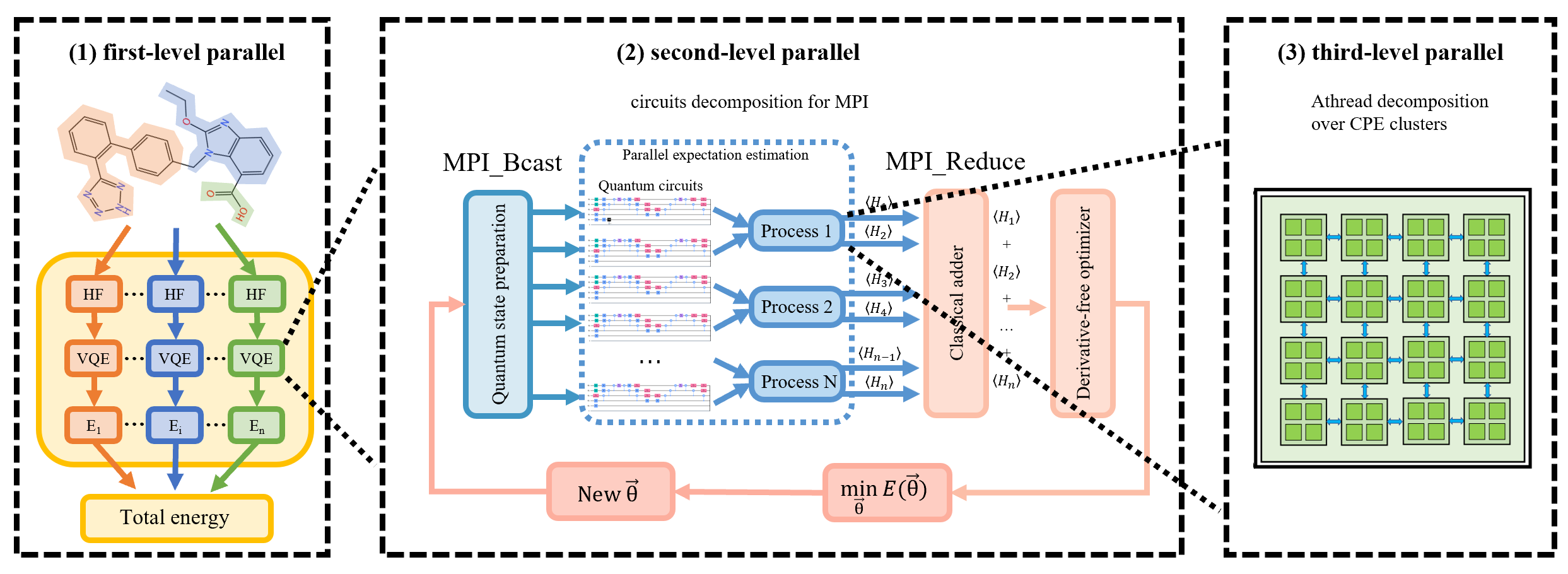}}
	\caption{Workflow with the Julia version of \system, including the task-level, the process-level and thread-level parallelization for the quantum chemistry simulation. }
	\label{fig:parallel}
\end{figure*}

Three levels of parallelization are used for the DMET-MPS-VQE simulator. As shown in Fig.~\ref{fig:parallel}, the calculation of different fragments can be performed in an embarrassingly parallel manner, that is, the first level of parallelization. As no communication between different fragment calculations is required,  we split the whole CPU pool into different sub-groups and sub-communicators.  Within each sub-group, the total energy is calculated with the VQE method. 
Within each VQE calculation, another two levels of parallelization have been used, that is, the second level over circuits and the third level over linear algebra operations. In the VQE calculation, the electronic molecular Hamiltonian $H$ can be written as a sum of 
the Pauli strings: $H=\sum_i{a_i P_i}$, where each Pauli string $P_i \in \{I, \sigma_x,\sigma_y,\sigma_z \}^{\otimes n} $ is a tensor product of $n$-qubit Pauli operators. We can estimate each $\langle\psi|P_i|\psi\rangle$ independently and then sum over all those results with corresponding coefficients ($a_i$) to get $\langle\psi|H|\psi\rangle$. This enables high parallel scalability with adapted dynamical load balancing algorithm. The parallel simulation algorithm based on distributed memory over the circuits, just ``mimic'' the actual quantum computers (for quantum computers each term $\langle\psi|P_i|\psi\rangle$ also has to be evaluated individually), so our method can offer a good reference for VQE running on the quantum computers.
For tensor operations including tensor contraction and singular value decomposition (SVD), a low-level multi-threaded parallelism on the CPE is used to further boost the performance. 

\subsection{The memory-efficient scheme to store the circuits}
\begin{figure}[t]
	\centering
	\includegraphics[width=0.8\linewidth]{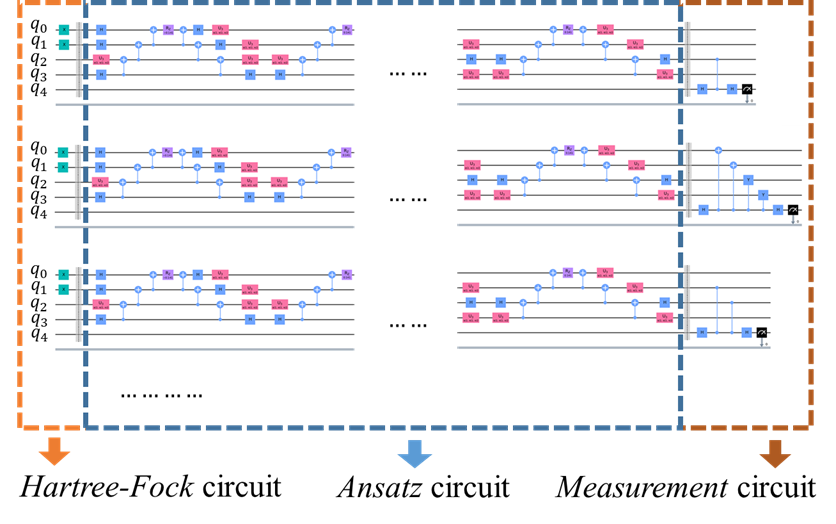}
	\caption{The first three circuits for simulating H$_2$ molecule using VQE. The experiment contains 15 circuits, corresponding to 15 Pauli terms from the H$_2$ Hamiltonian (under Jordan-Wigner transformation).}
	\label{fig:circ_h2}
\end{figure}

The second level of parallelism is based on the fact that the electronic structure Hamiltonian can be expressed as the summation of a polynomial number of mutually uncorrelated Pauli strings. Expectation values of each Pauli string can thus be calculated independently. Fig.\ref{fig:circ_h2} gives an example for the hydrogen molecule. The $4$-qubit Hamiltonian of hydrogen molecule contains $15$ Pauli strings, therefore $15$ circuits are required for each of them. For each circuit, the first $2$ $X$ gates prepare a \textit{Hartree-Fock} reference state. The next $120$ gates (with variational parameters in the $RZ$ gates) on q$_0$, q$_1$, q$_2$ and q$_3$ constitute the \textit{ansatz} circuit, which corresponds to the first-order Trotterized UCC ansatz and is identical for each Pauli string. $q_4$ is the ancillary qubit for Hadamard test used to compute the expectation value, and the gates on $q_4$ in different circuits corresponding to the \textit{measurement} part are constructed according to each Pauli string. As shown in Fig.\ref{fig:parallel}, during the VQE optimization, circuits corresponding to mutually exclusive subsets of Pauli strings in Eq.\ref{eq:qubit-ham} are mapped to different processes. The energy $E(\theta) = {\langle \psi(\theta) | \hat{H} | \psi(\theta) \rangle}$ under current parameters $\{\theta\}$ is obtained by reducing the results on different processes and then provided to the optimizer.

It should be noted that the number of Pauli strings in the electronic Hamiltonian scales as $\mathcal{O}(N_{q}^{4})$, where $N_{q}$ is the number of qubits. Even for a medium-size molecule such as benzene (72 qubits, 330816 Pauli strings), storing all the $330816$ circuits on each process will bring a lot of pressure on the memory space of CGs and synchronizing the circuits after each optimization step will also result in remarkable overhead on communication. Noticing that the ansatz part of different circuits are identical and the measurement part can be constructed on-the-fly in the first energy evaluation and then kept constant during the VQE optimization, a straightforward optimization for memory usage can be achieved by keeping only one replica of the ansatz circuit on each process. The speedup of such a memory-optimized simulation is given in Sec.~\ref{subsection:speedup}.

\subsection{Using Julia Programming Language on Sunway}
\label{sec: julia}
Julia is designed to take advantage of modern techniques for executing dynamic languages efficiently. It has the performance of a statically compiled language while providing interactive dynamic behavior and productivity\cite{bezanson2012julia}. The key ingredients of performance are:
\begin{itemize}
	\item Rich type information, provided naturally by multiple dispatch;
	\item Aggressive code specialization against run-time types;
	\item JIT compilation using the LLVM compiler framework\cite{lattner2004llvm};
\end{itemize}

We build the core components of  \system ~based on the Julia language to take advantage of the above advantages. Julia codes can be highly extensible thanks to its type system and multiple dispatch mechanism, and the meta-programming ability makes developing customized syntax and device-specific programs simple\cite{2019Yao}. Generic programming in Julia helps us to reach optimized performance while still keeping the code general and concise. Julia integrates well with other programming languages. It is generally straightforward to use external libraries written in other languages in \system. We have also a python version of \system, which will be reported elsewhere.  In this work, the electronic structure code PySCF\cite{openfermion} and openfermion\cite{pyscf1} are used to calculate chemical quantities such as the orbital occupations and qubit Hamiltonian for quantum state evolution and measurements.
\begin{figure}
	\centerline{\includegraphics[width=0.95\columnwidth]{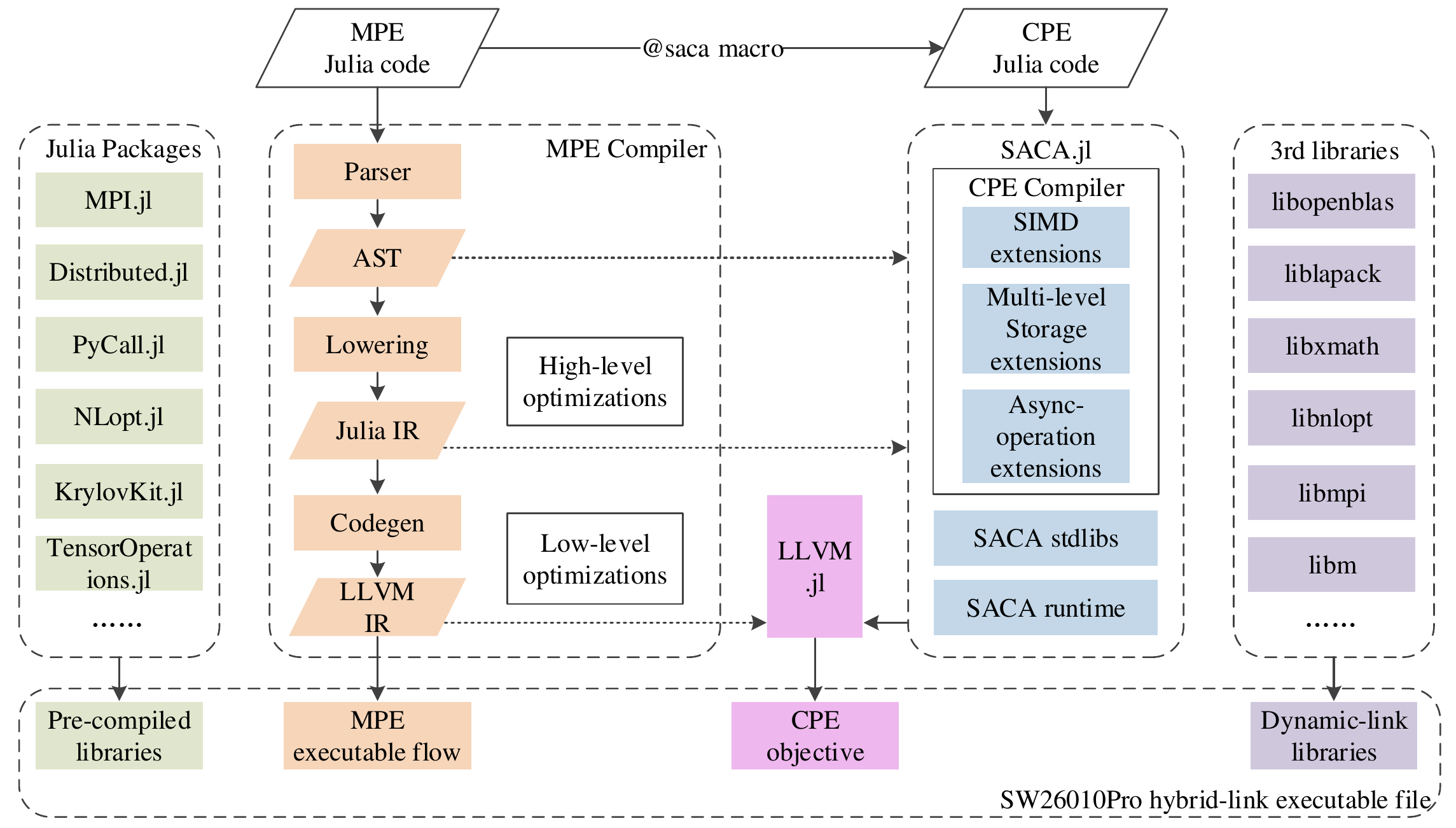}}
	\caption{Schematic overview of the Julia compiler on the Sunway heterogeneous many-core processor. Dashed arrows indicate meta-programming interfaces, while solid arrows represent the flow of code. }
	\label{fig:swjulia}
\end{figure}

We unleash the Julia compiler on the new Sunway supercomputer and deploy \system ~on it. Fig.~\ref{fig:swjulia} shows the schematic overview of our Sunway Julia compiler. The MPE compiler is the main part which serves as a JIT compiler for MPEs. The SACA.jl library focuses on providing definitions for CPE operations that are required for writing effective applications based on the SACA programming model\cite{li2021sw_qsim}. The main interface for calling functions on CPEs resembles a call to an ordinary Julia function: \texttt{@saca (config...) function(args...)}, where the \texttt{config} tuple indicates the spawn configuration similar to the arguments in SACA APIs. In addition, Julia features powerful meta-programming and reflection capabilities. For example, the high-level Julia IR is accessible with the \textsl{code\_lowered} and \textsl{code\_typede} reflections, and can be modified with generated functions. The CPE Julia Compiler implements multiple types of extension based on SW26010Pro architecture features. The SIMD extensions support the definition of non-uniform width short vector data types and the built-in operations in Julia language layer. The multi-level storage extensions make it easy for CPE Julia code to migrate date between multi-level storage, such as LDM, TLS, CROSS, etc. The async-operation extensions provide flexible interfaces for DMA, RMA and other asynchronous components. To facilitate interactions with LLVM, we adopt LLVM.jl\cite{LLVM.jl-2017} package which provides a high-level wrapper to the LLVM C API, using Julia’s powerful Foreign Function Interface (FFI) to interact efficiently with the underlying libraries. To meet the application requirements of  \system, we port a number of Julia standard and third-party libraries, such as MPI.jl\cite{byrne2021mpi}, Distributed.jl, PyCall.jl, etc. MPI.jl and Distributed.jl are two commonly used distribution modes of Julia. Distributed.jl is the method to spawn new processes in separate memory spaces. Like how multi-threading is set up, its master-slave distributed architecture makes it nice and convenient for smallish parallelization. MPI.jl is a basic Julia wrapper for the Message Passing Interface (MPI)\cite{gropp1999using}. The MPI libraries are versatile and highly optimized for SW26010Pro, so that the MPI.jl could be easy to use it from Julia by preparing functions with almost the same name as C functions and get similar performance. Given the performance and scalability of \system ~on the new Sunway supercomputer, we finally chose MPI.jl as the main way for distributed computing. We use swBLAS as the backend for matrix-vector and matrix-matrix multiplications. PyCall.jl is used to directly call python packages from Julia.

\change{
When compiling \system, each computing node has a separate JIT compilation engine, and there is no data interaction during JIT compilation, so it can be easily scaled up to more than 20M cores. For more than 20M cores, the simultaneous loading of the Julia dynamic libraries on computing nodes can become a performance bottleneck. We avoid this problem by using the ROFS (Read Only File System). Compared with the GFS (Global File System), The ROFS has smaller capacity but faster access speed, which significantly reduces the loading time of Julia dynamic libraries.
}

When running \system, the hotspots are mainly tensor contraction and the SVD functions in the Julia's basic library LinearAlgebra.
In the tensor contraction,  the first step is  the index permutation of the tensors, followed by the matrix multiplication~(ZGEMM) to accomplish the calculation. Here we use the fused permutation and multiplication technique. In the SVD calculation, the matrix is transformed into a bidiagonal matrix  using an orthogonal transformation, then the bidiagonal matrix is diagonalized using  BDC (bidiagonal divide-and-conquer) or QR decomposition. Thus the core calculations of SVD are mainly the matrix vector multiplication, matrix multiplication, vector multiplication and matrix transpose, which is done with the swBLAS library.


\change{We also note that running \system ~on non-Sunway architectures needs to address the following issues:
\begin{enumerate}
\item	The compiling and running environment for Julia program on computing nodes: The Julia compilation framework needs to be supported on non-Sunway architectures to meet the compilation and running requirements of the Julia language. Currently the latest Julia  version only supports i686 / x86\_64 / arm / aarch64 / powerpc64le. Other architectures require porting of the Julia compilation framework;
\item	MPI.jl distributed environment: The underlying MPI protocol and the MPI.jl library need to be supported on non-Sunway architectures to meet \system's massively scalable communication requirements. All of the mainstream supercomputer systems support the MPI protocol, and some architectures need to transplant the MPI.jl library;
\item	Architecture specific optimization: Optimized third-party computing libraries, including BLAS and LAPACK, need to be supported on non-Sunway architectures to meet \system's computing performance requirements. The current mainstream non-Sunway Architectures basically meet this requirement.
\end{enumerate}
Therefore, in general \system ~has good portability and can be rapidly and efficiently transplanted to the current mainstream non-Sunway architectures.
}

\section{EVALUATION}
\subsection{Simulation Validation}
\begin{figure}
	\centerline{\includegraphics[width=0.95\columnwidth]{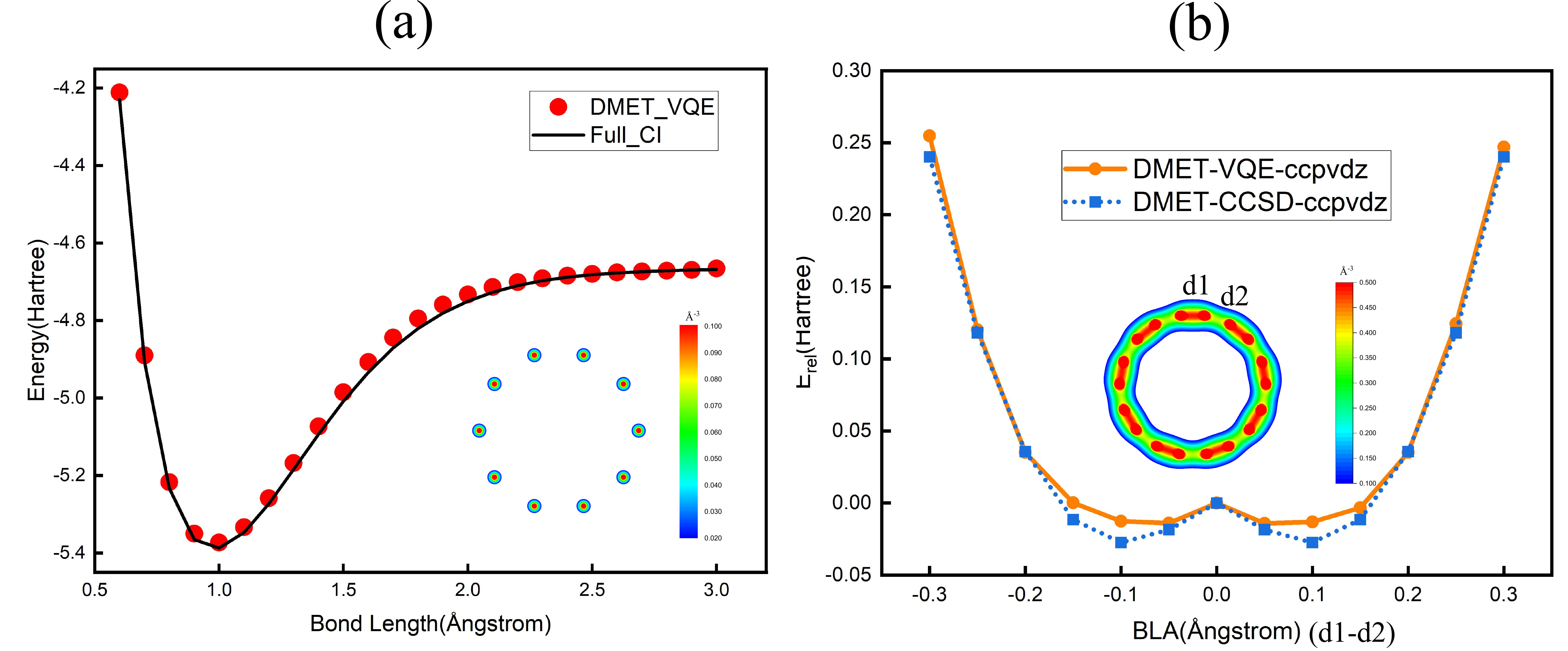}}
	\caption{(a) The accuracy of our \system  ~method for hydrogen ring composed of 10 atoms and comparing with full CI results. (b) The potential energy curve of the C18 molecule with cc-pVDZ basis set.}
	\label{fig:validation}
\end{figure}

We firstly validate our method by comparing with the FCI results as shown in Fig.~\ref{fig:validation}(a). We start with the calculation of the hydrogen ring with 10 atoms to show the effectiveness of the DMET-MPS-VQE method. Restricted Hartree-Fock method is used as the low-level calculation of the whole system, with the STO-3G basis. In the DMET-MPS-VQE calculation, the hydrogen atoms are divided into fragments with two atoms. Here we can see the potential energy curve with the DMET-MPS-VQE method,  which shows very good agreement with those from FCI, and the relative errors are within 0.5\%. We also compare the simulation results of MPS-VQE with FCI, and the results of the ground state energy are very accurate for the chemical systems of H$_2$, LiH, and H$_2$O, with the relative errors at the level of 0.01\%.

We further examine the carbon ring molecule (C18) with a large basis set (cc-pVDZ) as shown in Fig.~\ref{fig:validation}(b). The equilibrium geometry of this molecule is the bond-length alternated structure as confirmed in experiment\cite{Kaiser2019}. Here we scan the energy of this molecule with the change of the  bond length alternation (BLA), defined as the lengths of the two sets of C-C bonds in the molecule. We show results with cc-pVDZ basis set~(for carbon, the 1s orbital is frozen, only the 2s and 2p basis orbitals of carbon atom are considered in the VQE calculation, while the high angular momentum orbitals are treated at the mean-field level). We can see that, similar to the CCSD results, our DMET-VQE results predict that the energy of the bond-length alternating structure is lower, which in good agreement with experimental observations.

\subsection{Speedup}
\label{subsection:speedup}
\begin{figure}
	\includegraphics[width=0.95\columnwidth]{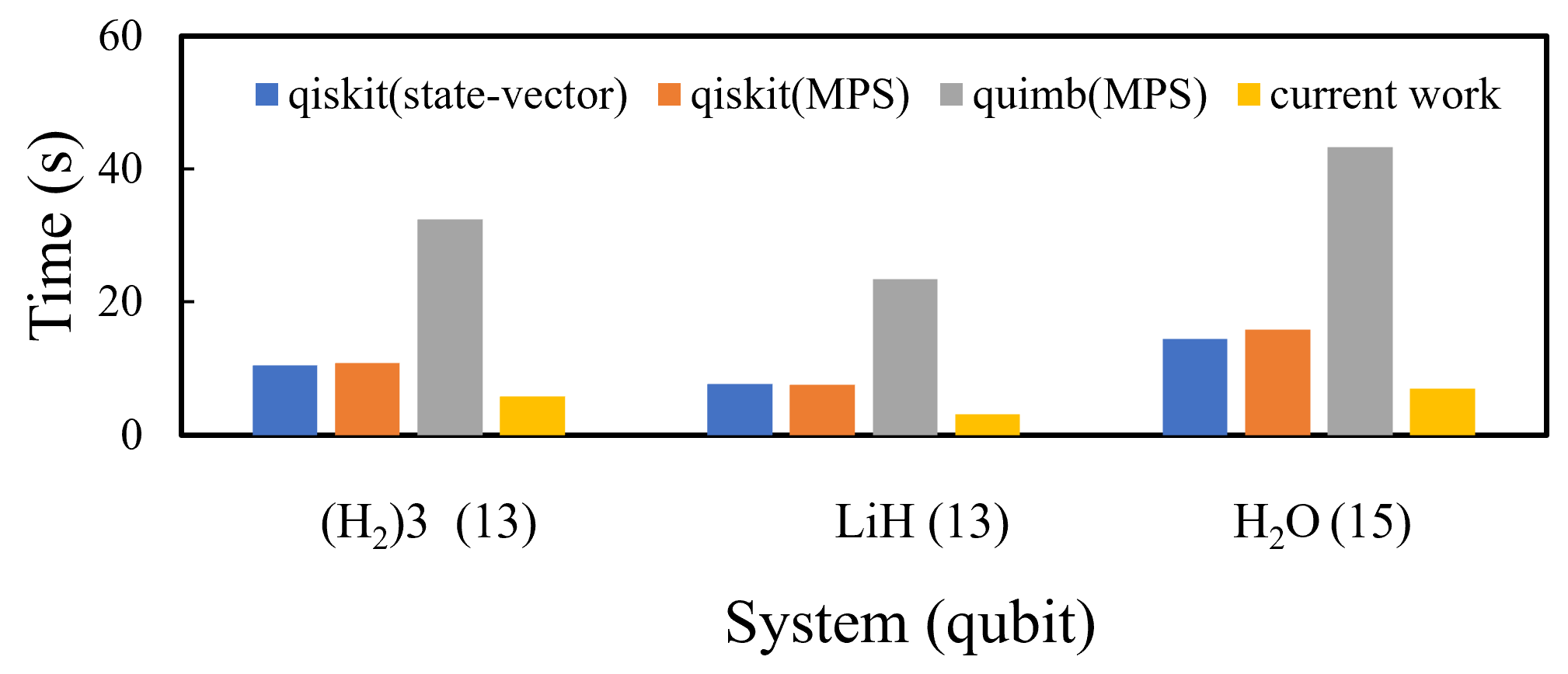}
	\caption{ The performance comparison between different software for the quantum simulation of the real chemical systems. Here the current work refers to \system.}
	\label{fig:software_compare}
\end{figure}

In this section, we compare two other libraries with our proposed algorithm and analyze the test results. The two libraries are qiskit (using the state vector and MPS algorithm) and quimb (using the MPS algorithm).  We evaluate the four methods by comparing the simulating time of one circuit for three molecule systems using one process, as shown in Fig.~\ref{fig:software_compare}. The computing time of quimb~(MPS) is on average $3$ times slower than qiskit (state-vector). Our method (\system) is on average around $7$ times faster than quimb and around $2$ times faster than qiskit~(both MPS and state vector). 

%

\begin{figure}
	\includegraphics[width=0.95\columnwidth]{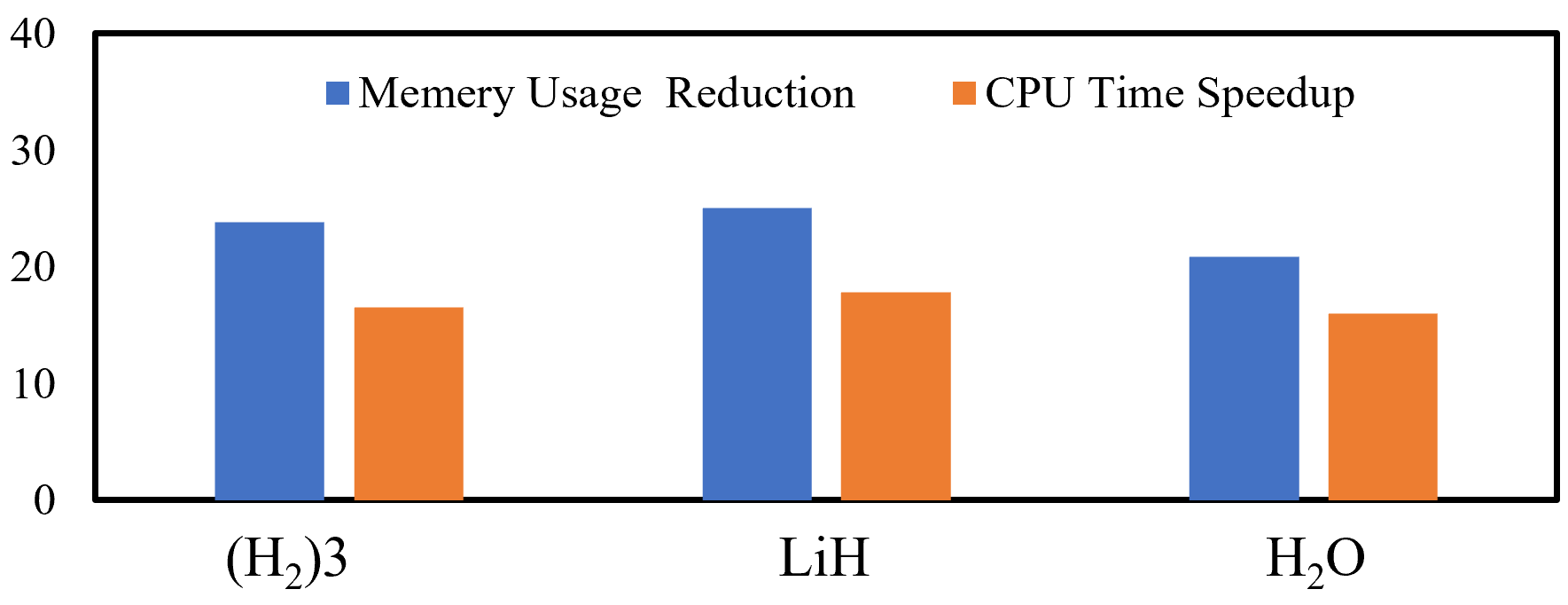}
	\caption{ The performance comparison with the memory-efficient scheme in the quantum simulation of the real chemical systems. }
	\label{fig:mem_opt}
\end{figure}

\begin{figure}
	\centering
	\includegraphics[width=0.98\columnwidth]{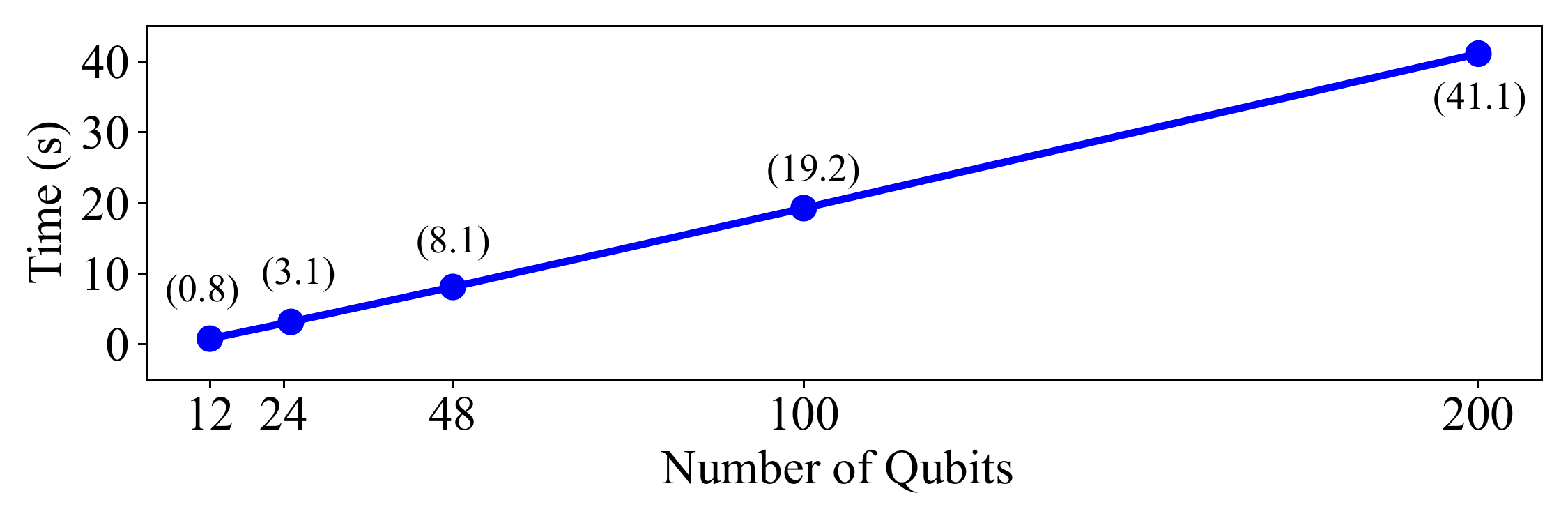}
	\caption{ The simulation time~(annotated in parentheses) of one VQE circuit with our MPS-VQE simulator for hydrogen chain molecules containing 6 to 100 atoms (number of qubits changes from 12 to 200). }
	\label{fig:mps_vqe_test}
\end{figure}

Figure~\ref{fig:mem_opt} shows the result for the memory-efficient optimisation in the circuit simulation. Here we choose three systems as examples. The (H$_2$)3, LiH and H$_2$O molecules have $919$, $630$ and $1085$ circuits respectively. The number of circuits per process is $18$, $19$, $17$. We can see the speedup is around 15x and the memory reduction is around 20x. This indicates that the memory-optimization scheme is effective and efficient.

Figure~\ref{fig:mps_vqe_test} shows the simulation time with the MPS-VQE simulator for the VQE circuits of the hydrogen chain. The number of the 
electrons/atoms changes from 6 to 100, while
the corresponding number of the qubits changes from 12 to 200. It is clearly shown that the computing time scales linearly with the number of qubits, which demonstrates the power of \system. 
\change{There are two major bottlenecks for further scaling up to larger systems: 1) the memory requirement, which grows linearly with the number of qubits ; 2) the quantum circuit corresponding to more complex systems could be significantly deeper, for which one needs a larger bond dimension.}

\begin{figure}
	\includegraphics[width=1.0\columnwidth]{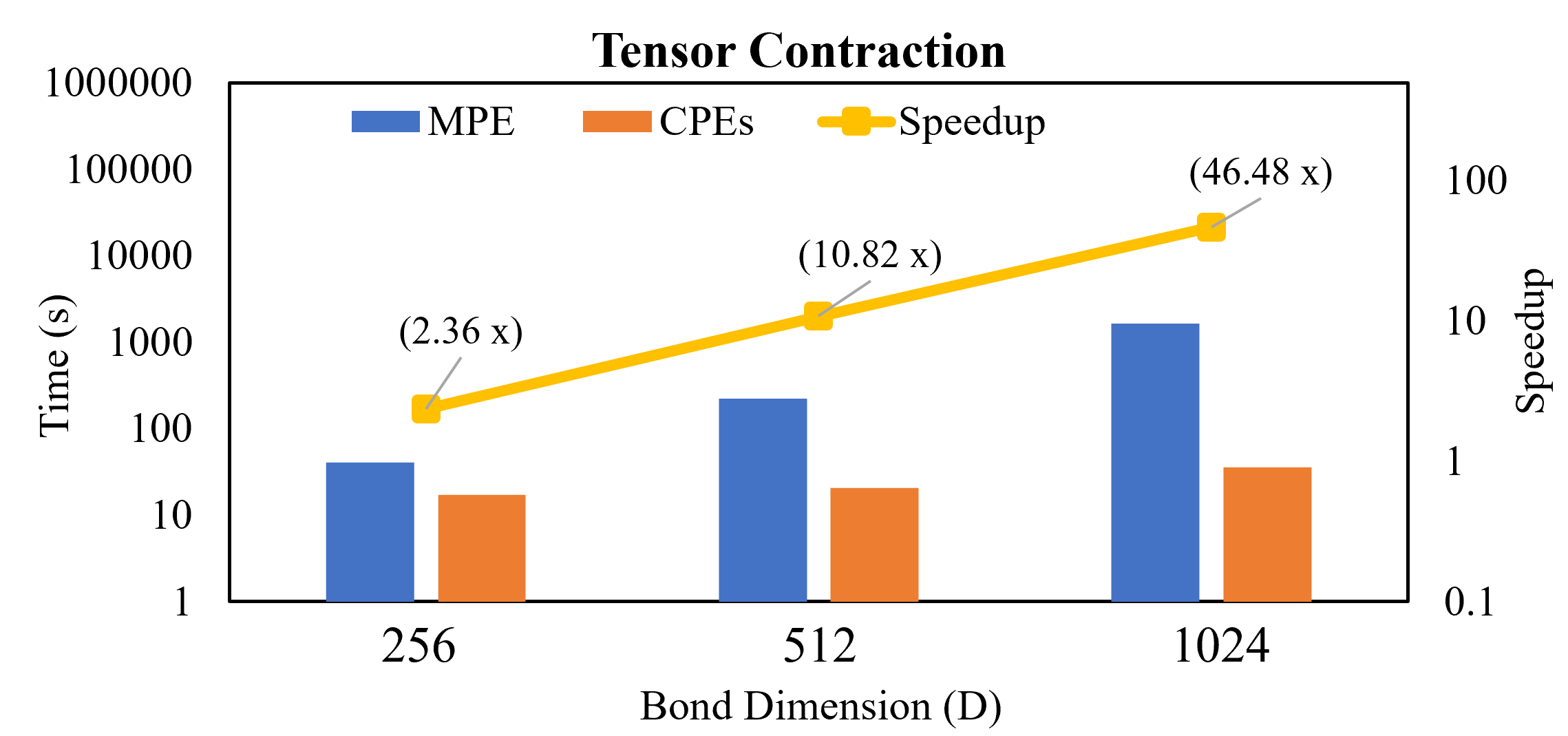}
	\includegraphics[width=1.0\columnwidth]{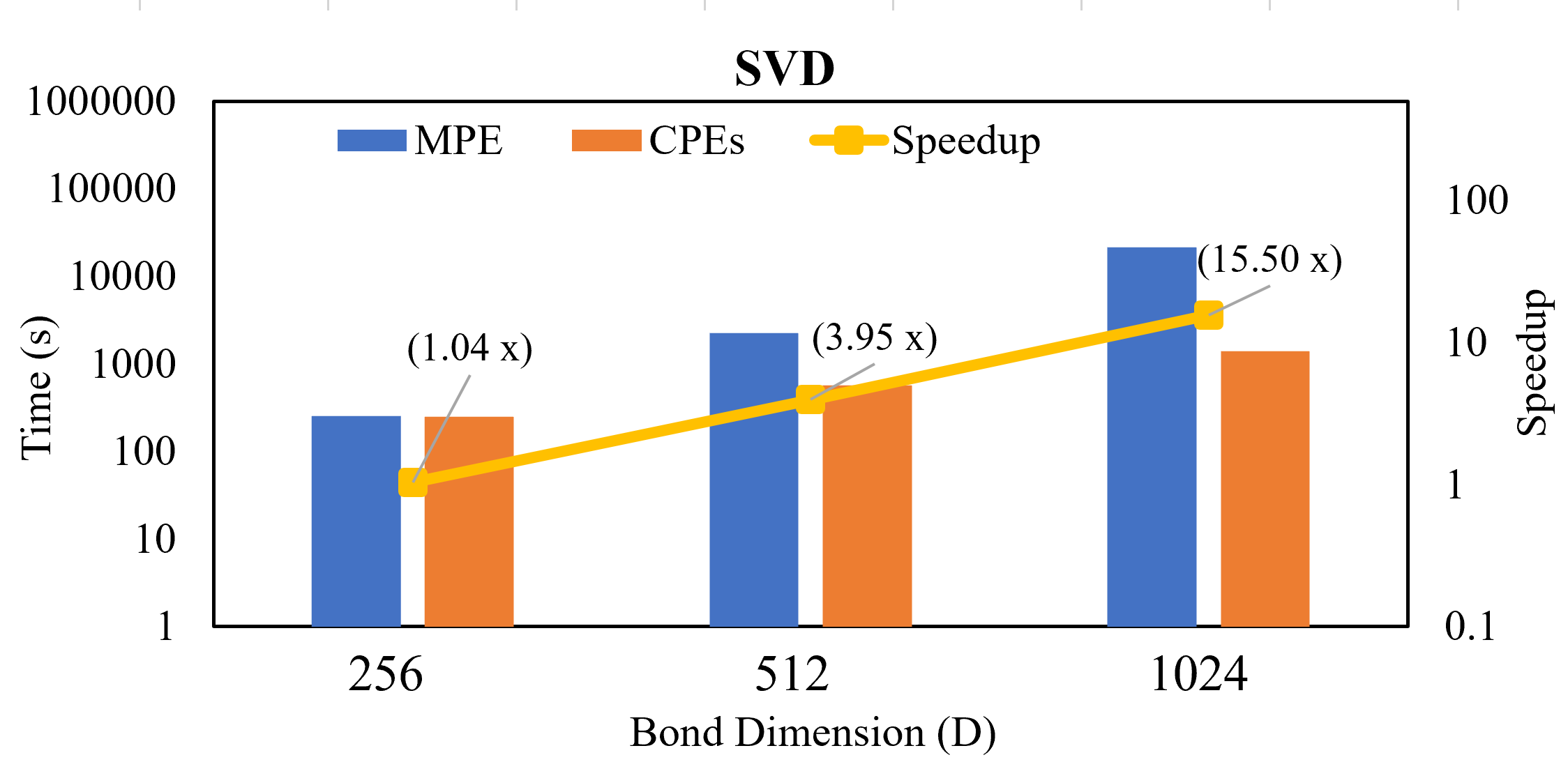}
	\caption{The performance comparison with respect to the bond dimension. The upper figure refers to the performance of the tensor contraction and the lower figure refers to the performance of SVD. }
	\label{fig:svd_speedup}
\end{figure}

In the MPS-VQE simulation, the percentages of time spent in tensor contraction and SVD are around 15\% and 82\% when the number of qubits changes from 33 to 129. Fig.~\ref{fig:svd_speedup} give the evaluation of the tensor contraction and SVD performance improvements when using different bond dimensions. We observe that compared to the original version that only uses the MPE, the optimized version use both the MPE and the 64 CPEs produces an overall speedup in the tensor contraction calculation ranging from 2.3x to 46.5x and the SVD calculation ranging from 1.04× to 15.5×, and the speedup is more significant when increasing the bond dimension \change{from 256 to 1024}.

\change{We also evaluate the performance of \system ~on the x86 (AMD Ryzen EPYC 7452) CPU, using a quantum circuit similar to that used in Fig.~\ref{fig:mps}(c), with unitaries replaced by 2-qubit gates acting on neighbouring qubits. The initial quantum state is generated randomly according to a bond dimension threshold (which is 512 in this test). We observe that the SW version is approximately $1.1$ times faster than the x86 version based on OpenBLAS and 16.6 times faster than the x86 version based on LAPACK-3.2.}


\subsection{Scalability Results}

\begin{figure}
	\centering
	\includegraphics[width=0.95\linewidth]{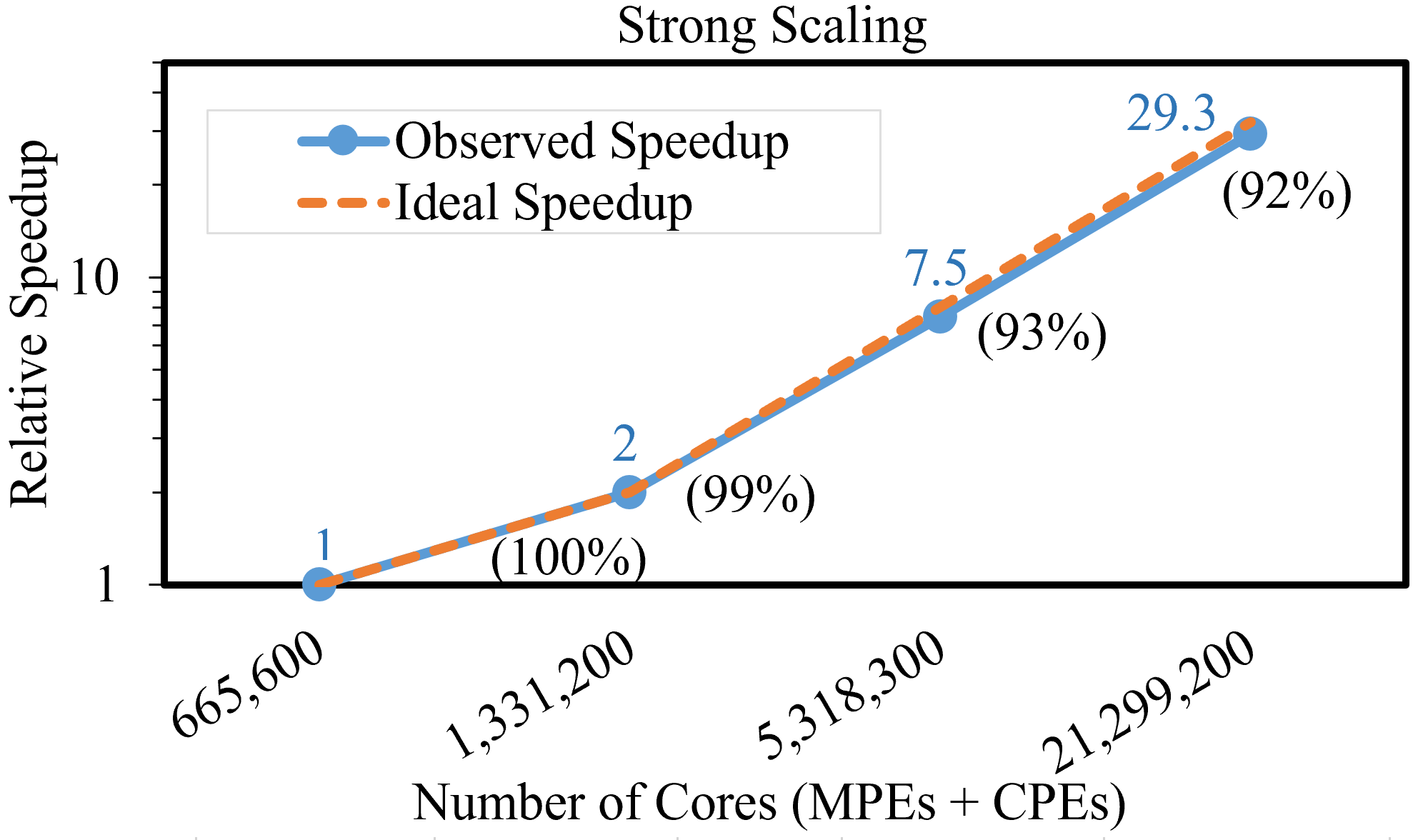}
	\caption{Strong scalability of the quantum computation time for the hydrogen chains. The blue line is the relative speedup and the orange dash line is the ideal speedup. Parallel efficiency values are annotated in  parentheses, respectively. }
	\label{fig:stronge}
\end{figure}

Figure~\ref{fig:stronge} shows the strong scaling performance of the quantum computational chemistry simulation for the hydrogen chain with 1280 atoms~(1280 electrons). In all calculations, 
each MPI sub-group is mapped to 2048 processes. The code shows good strong scaling performance: as the number of Sunway processes increases from $10,240$ to $327,680$ for the calculation (the number of cores increases from $665,600$ to $21,299,200$), the parallelization efficiency exceeds 92\% and 30$\times$ speedup with respect to the $10,240$ processes is achieved.

\begin{figure}
		\centering
	\includegraphics[width=0.99\linewidth]{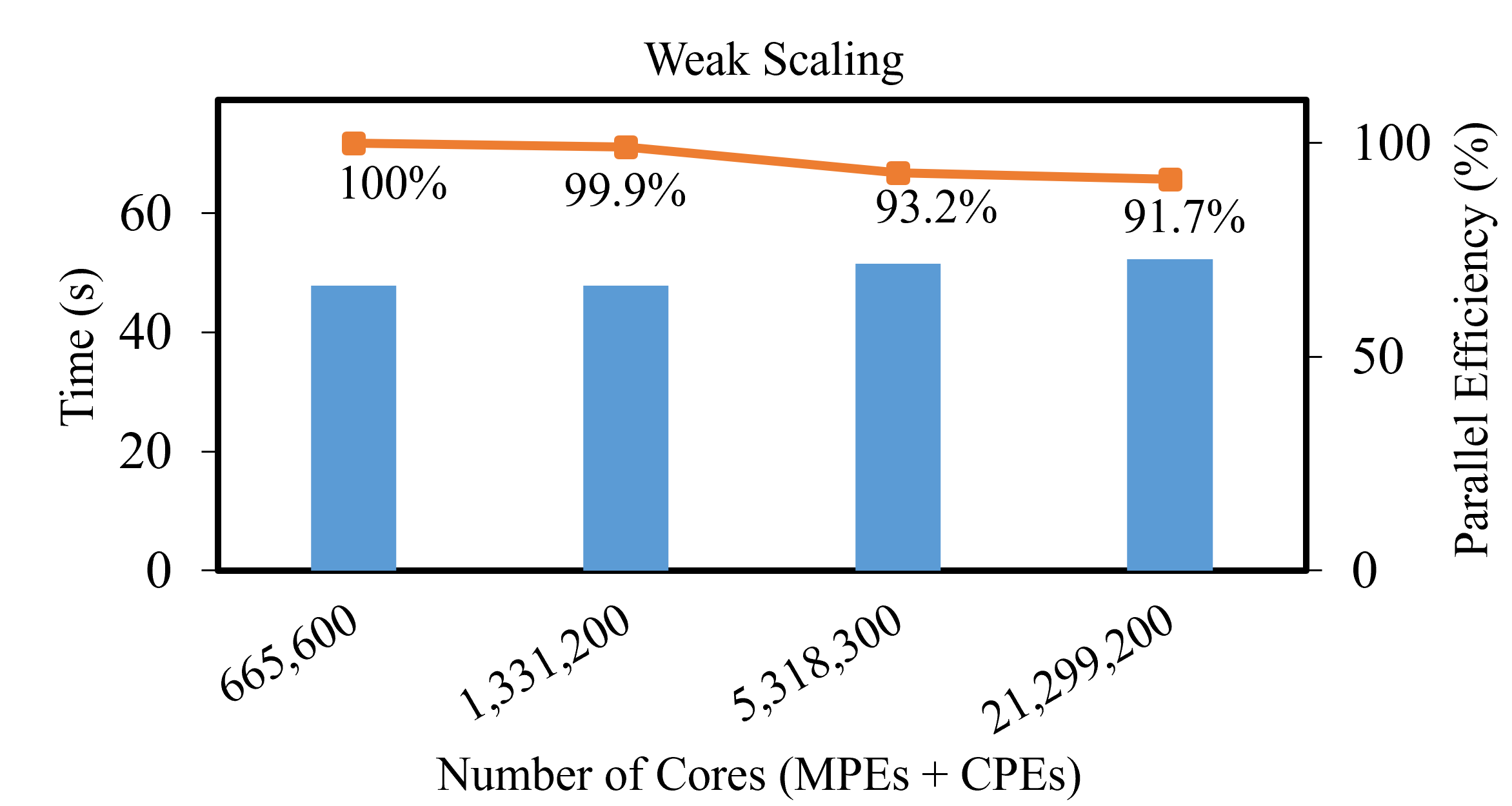}
	\caption{The weak scalability for the quantum simulation time of the hydrogen chain system. Here the number of the atoms in the simulation calculation are increased with the number of the cores (MPEs + CPEs). The blue bars is the simulation time, the orange line is the weak parallel efficiency.}  
\label{fig:weak}
\end{figure}

Fig.~\ref{fig:weak} shows the weak scaling of the quantum computational chemistry simulation for the hydrogen chain. To obtain the weak scaling data, we time the MPS-VQE calculations with an increasing number of fragments~($2048$ processes per fragment). The number of atoms~(electrons) are (40, 80, 320, 1280) respectively in the weak scaling calculation.  The calculation achieves excellent weak scaling, exhibiting high parallelization efficiency as the number of cores grows. Taking $10,240$ processes 
as a reference, the code achieves a parallelization efficiency of around 92\% with $327,680$ processes~(  $21,299,200$ cores). 

\change{The reason for the high parallelization efficiency is due to the low data communication overhead in our algorithm. There is almost no data communication on the DMET level parallelization except the final summation over all fragments (a scalar from each process) to get the total DMET energy. On the MPS-VQE level parallelization over 2048 processes, we first need to broadcast the parameters from the root process to all processes using MPI\_Bcast, and after the computation, we need to collect all the results to the root process using MPI\_Reduce, which are shown in Fig.~\ref{fig:parallel}. We use the Julia performance profiler to measure the communication costs, which shows that the amount of data communication per process is around $15.6$ KB and the time spent on data communication is less than $0.001$ second per VQE iteration.}


\section{Implications}
\begin{figure}
\centerline{\includegraphics[width=0.93
	\columnwidth]{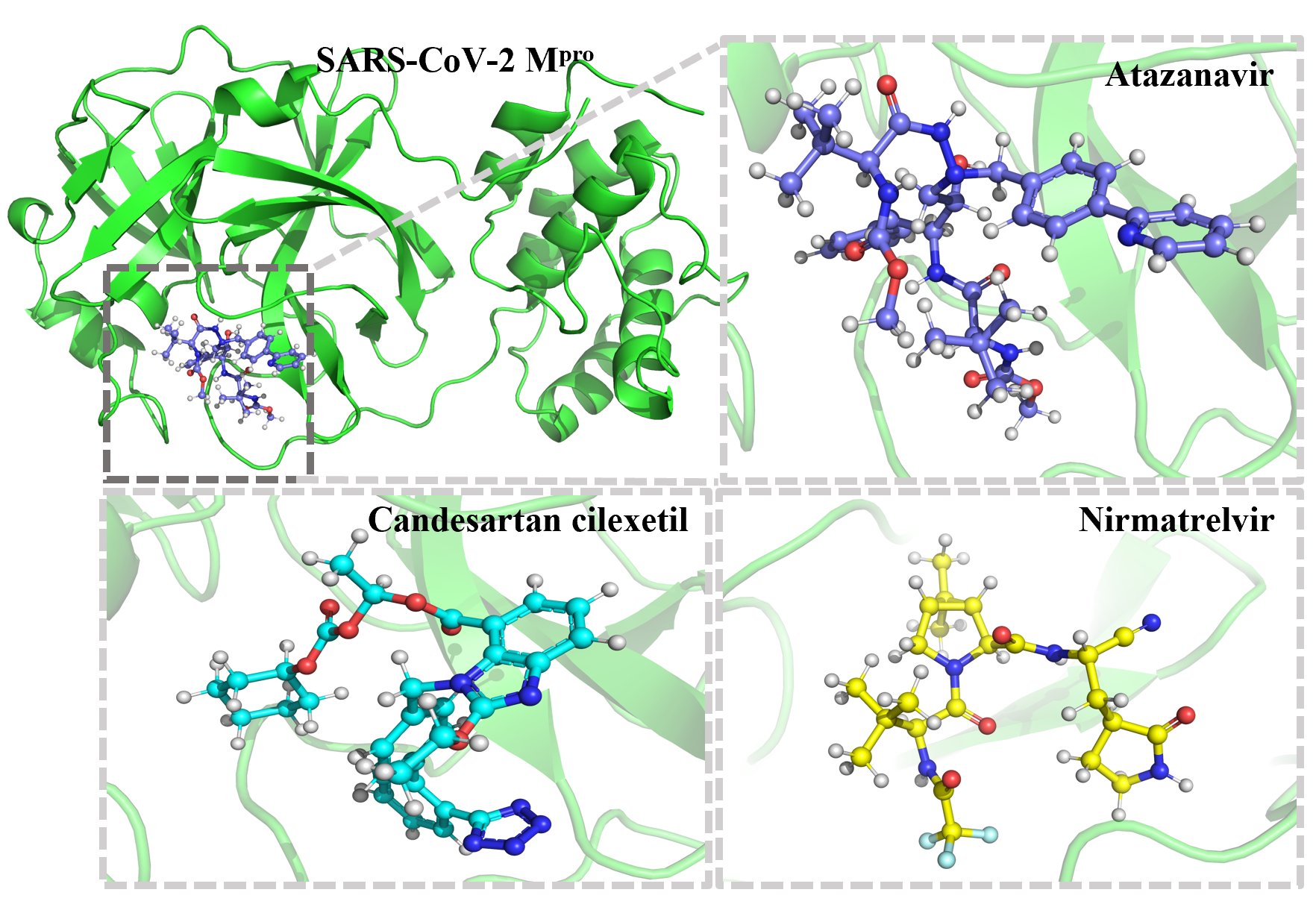}}
\caption{The SARS-CoV-2 M$^{\rm pro}$ protein with a ligand~(Atazanavir) bound to its active site and the molecular structures of 3 confirmed inhibitors. }
\label{fig:mpro_3_ligands}
\end{figure}

Here we apply our \system ~method in real chemical systems for the quantification of protein-ligand interactions. Compared to the empirical force fields method, the quantum mechanical calculations can automatically include effects of polarization, charge transfer, charge penetration, and the coupling of the various terms. Therefore, the current quantum mechanical approaches could offer more accurate and detailed information on the nature of protein–ligand interaction, which is valuable in high-accuracy binding affinity prediction and so does in drug design. The SARS-CoV-2 main protease~(M$^{\rm pro}$) is an enzyme that cleaves the viral polyproteins into individual proteins required for viral replication, so it is the important target to develop drugs for SARS-CoV-2. The structure of the M$^{\rm pro}$ complex originates from Ref.\cite{Liu2020}(PDB 6lu7), as shown in Fig.~\ref{fig:mpro_3_ligands}. Here by using the "frozen protein" approximation\cite{Kirsopp2021} to remove the protein-protein interactions, we use
$E_{\rm b}=E_{\rm ligand\_in\_protein}-E_{\rm ligand}$ to get the binding energy $E_{\rm b}$, which is then used as the metric for ranking. 
We use the geometries of the 13 neutral ligands from Ref.\cite{Wang2021}, which has been optimized with density functional theory method. Then we use \system ~method to get the $E_{\rm b}$, the geometries of $E_{\rm ligand}$  are optimized  at Hartree-Fock level to account for some amount of the geometric distortion needed for the ligand to occupy the active site. We examine the binding energy of 13 ligands~(The largest molecule is Atazanavir contains 103 atoms and 378 electrons), we find Candesartan cilexetil binds best~($E_{\rm b}$ is -6.8 eV), which is in good agreement with the experimental observation~\cite{LiZhe2020}. 
We also examine the Nirmatrelvir within Paxlovid (Pfizer’s novel anti-viral Covid-19 drug), 
which is an orally bioavailable protease inhibitor that is active against M$^{\rm pro}$, the binding energy $E_{\rm b}$ is -7.3 eV, even lower than that of Candesartan cilexetil, which indicates that the performance of Nirmatrelvir is better than Candesartan cilexetil. 

\section{CONCLUSION}
The innovations realized in this study demonstrate that \system ~is suitable for large-scale simulation of quantum computational chemistry, based on a combination of the Density Matrix Embedding Theory and the Matrix Product States to reduce the exponentially memory scaling against the system size;  a customized three-level parallelization scheme has been implemented  according to the nature of the physical problem and the many-core architecture; Julia is used as the primary language which both makes programming easier and enables cutting edge performance close to native C or Fortran; Real chemical systems have been studied to demonstrate the power of \system ~in computational quantification of protein-ligand interactions. To the best of our knowledge, this is the first reported quantum computational chemistry simulation calculation for real chemical system with as many as $100$ atoms and $1000$ qubits \change{using DMET-MPS-VQE (and $200$ qubits using MPS-VQE)}, and scales to around $20$ million cores.  
This paves the way for benchmarking with near term VQE experiments on quantum computers with around $100$ qubits. 

\change{In future work we will also attempt to port \system ~to other computing architectures, for which the following issues need to be addressed: First, the Julia compilation framework needs to be supported to meet the compilation and running requirements of the Julia language; Second, the underlying MPI protocol and the MPI.jl library need to be supported to meet \system's massively scalable communication requirements; Finally, optimized third-party computing libraries, including BLAS and LAPACK, need to be supported to meet \system's computing performance requirements.}

\bibliographystyle{./IEEEtran}
\bibliography{./Manuscript.bib}

\end{document}